\documentclass[a4paper,11pt]{article}

\usepackage{jcappub_nt}
\usepackage[utf8]{inputenc}

\usepackage{bm}

\title{Doubly coupled matter fields in massive bigravity}

\author[a,b]{Xian Gao}

\author[c]{and Lavinia Heisenberg}

\affiliation[a]{Research Center for the Early Universe (RESCEU), Graduate School of Science,\\
	The University of Tokyo, Tokyo 113-0033, Japan}

\affiliation[b]{Department of Physics, Tokyo Institute of Technology,\\
	2-12-1 Ookayama, Meguro, Tokyo 152-8551, Japan}

\affiliation[c]{Institute for Theoretical Studies, ETH Zurich,\\
	Clausiusstrasse 47, 8092 Zurich, Switzerland
}

\emailAdd{gao@resceu.s.u-tokyo.ac.jp}
\emailAdd{lavinia.heisenberg@eth-its.ethz.ch}

\abstract{In the context of massive (bi-)gravity non-minimal matter couplings have
been proposed. These couplings are special in the sense that they are free of the
Boulware-Deser ghost below the strong coupling scale and can be used consistently 
as an effective field theory. Furthermore, they enrich the phenomenology of massive
gravity. We consider these couplings in the framework of bimetric gravity and study
the cosmological implications for background and linear tensor, vector, and scalar perturbations. 
Previous works have investigated special branch of solutions. Here we perform a
complete perturbation analysis for the general background equations of motion
completing previous analysis.
	}

\keywords{cosmological perturbation theory, modified gravity}

\arxivnumber{}

\begin{document}
	\begin{flushright}
		RESCEU-22/16
	\end{flushright}
\maketitle


\section{Introduction}
The high precision cosmological observations made it possible to test the underlying fundamental theory of gravity. Together with the assumption of General Relativity (GR) as being the right theory and the cosmological principle, the universe is well described by the $\Lambda$CDM model. It constitutes a predominant amount of dark energy in form of a cosmological constant and dark matter. Aside from negligible reported anomalies  \cite{Ade:2015hxq}, the model is still the best fit to current cosmological data \cite{Ade:2015xua,Ade:2015lrj,Amendola:2012ys}. In spite of its observational triumph, the model suffers from serious theoretical problems, the most persisting one being the cosmological constant problem \cite{Weinberg:1988cp}.

An alternative scenario for the dark energy can be provided by infrared modifications of gravity. The simplest case corresponds to modifications in form of an additional scalar field \cite{Horndeski:1974wa,Nicolis:2008in,Deffayet:2009wt,Deffayet:2009mn,Deffayet:2011gz}. The presence of self-interactions of the scalar field and the non-minimal couplings to gravity yield interesting cosmological scenarios  \cite{DeFelice:2010pv,Kobayashi:2010cm,Kobayashi:2011nu,Gao:2011qe,Gao:2011mz,Gao:2011vs,DeFelice:2011uc,deRham:2011by,Heisenberg:2014kea}. Other interesting dark energy scenarios can be accommodated by considering a vector field as an additional field. The question about the consistent self-interactions of the vector field or similarly its non-minimal coupling to gravity is receiving a renewed interest lately  \cite{Horndeski:1976gi,EspositoFarese:2009aj,Jimenez:2009py,BeltranJimenez:2013fca,Jimenez:2013qsa,Jimenez:2014rna,Heisenberg:2014rta,Tasinato:2014eka,Allys:2015sht}.

It is an unavoidable question to pursue whether the graviton could be massive which would correspond to a natural infrared modification of gravity since the mediated force by a massive graviton would be suppressed on large scales. The weakening of graviton could be put on equal footing with recent cosmological acceleration.
At the linear level the theory is described by the Fierz and Pauli mass term \cite{Fierz:1939ix} without introducing ghostly sixth mode. This linear model however suffers from the vDVZ discontinuity \cite{vanDam:1970vg,Zakharov:1970cc} when the mass of the graviton is sent to zero since General Relativity is not recovered in that limit.
Actually, very soon after that Vainshtein realized that the linear approximation breaks down at some distance far from the source and that non-linear interactions become appreciable close to the source \cite{Vainshtein:1972sx}. Usually, these non-linear interactions reintroduce the ghostly six mode, the Boulware-Deser ghost \cite{Boulware:1973my}, and it was a challenging task to construct these potential interactions which would propagate only five physical degrees of freedom  \cite{deRham:2010ik,deRham:2010kj,Hassan:2011vm,Hassan:2011hr,Hassan:2011zd}. This ghost free theory of massive gravity is also technically natural and does not obtain strong renormalization by quantum corrections \cite{deRham:2012ew,deRham:2013qqa}. 

In the context of quantum stability of the theory, new ways of coupling the matter fields have been explored \cite{deRham:2014naa,Noller:2014sta,Heisenberg:2014rka}. The classical potential interactions had to be tuned in a very specific way to maintain the Boulware-Deser ghost absent and if one wants to keep this property also at the quantum level, only very restricted matter couplings through an effective composite metric are allowed. This effective metric is built out of the two metrics in such a way that the matter quantum loops would only introduce a running of the cosmological constant for the effective metric which in other words correspond exactly to the allowed potential interactions. This doubly coupled matter fields introduce already at the classical level the Boulware-Deser ghost\cite{deRham:2014naa, deRham:2014fha}, but the coupling through the effective metric is special in the sense that the decoupling limit of the theory below the strong coupling scale is maintained ghost free \cite{Huang:2015yga,Heisenberg:2015iqa}. Therefore, this coupling can be used as a consistent effective field theory. In the unconstrained vielbein formulation of the theory one can construct yet other type of effective metrics to which the matter fields can couple as well and the decoupling limit would still be free of the Boulware-Deser ghost \cite{Melville:2015dba}. Actually, the hope using the unconstrained vielbein formulation was to preserve the ghost freedom fully non-linearly with the original effective vielbein \cite{Hinterbichler:2015yaa}. Unfortunately, this resulted in a negative result and also in this formulation the Boulware-Deser ghost is reintroduced \cite{deRham:2015cha}. However, it is worth mentioning that if one is willing to break the local Lorentz symmetry, one can indeed achieve this fully non-linearly \cite{DeFelice:2015yha}. The inclusion of the doubly coupled matter fields has very important implications for cosmological applications \cite{deRham:2014naa,Enander:2014xga,Gumrukcuoglu:2014xba,Solomon:2014iwa,Gao:2014xaa,Comelli:2015pua,Gumrukcuoglu:2015nua,Lagos:2015sya,Heisenberg:2015wja,Heisenberg:2016spl,Heisenberg:2016dkj} as well as for dark matter phenomenology \cite{Blanchet:2015sra,Blanchet:2015bia,Bernard:2015gwa}.

The analysis of cosmological perturbations of the doubly coupled matter fields in massive gravity revealed that ghost and gradient instabilities can be successfully avoided together with the strong coupling issues since the vector and scalar perturbations maintain their kinetic terms \cite{Gumrukcuoglu:2014xba}. The application to the massive bimetric gravity yielded gradient instability in the vector sector and ghost instability in the scalar sector for one of the branch of solutions, whereas the other branch of solutions was free of any ghost instability. It is still an open question whether this second branch of solutions is also free from any gradient instabilities. The main purpose of the present work is to investigate the perturbation analysis of the bimetric gravity theory in the presence of the doubly coupled matter fields on top of a general background equations of motion, without specifying the branch and providing also the full quadratic action for the scalar perturbations. Thus, our work complete the analysis started in \cite{Gumrukcuoglu:2015nua}.

\section{Dynamical composite metric} \label{sec:dyncom}

A consistent coupling of some extra scalar field $\phi$ to both metrics simultaneously was introduced in \cite{deRham:2014naa} through a composite metric $\tilde{g}_{\mu\nu}$
	\begin{equation}
		\tilde{g}_{\mu\nu}\equiv\alpha^{2}g_{\mu\nu}+2\alpha\beta\,g_{\mu\lambda}X_{\phantom{\lambda}\nu}^{\lambda}+\beta^{2}f_{\mu\nu},
	\end{equation}
with $X_{\phantom{\mu}\nu}^{\mu}$ defined by
	\begin{equation}
		X_{\phantom{\mu}\lambda}^{\mu}X_{\phantom{\lambda}\nu}^{\lambda}\equiv g^{\mu\lambda}f_{\lambda\nu}.
	\end{equation}

We consider the same action as in \cite{Gumrukcuoglu:2015nua}
	\begin{equation}
	S=S^{g}+S^{f}+S^{\mathrm{pot}}+S^{\mathrm{com}},\label{action}
	\end{equation}
with
\begin{eqnarray}
S^{g} & = & \int\!\mathrm{d}^{4}x\sqrt{-g}\left(\frac{M_{g}^{2}}{2}R\left[g\right]+\mathcal{L}^{\mathrm{matter}}\left[g\right]\right),\label{S_g}\\
S^{f} & = & \int\!\mathrm{d}^{4}x\sqrt{-f}\left(\frac{M_{f}^{2}}{2}R\left[f\right]+\mathcal{L}^{\mathrm{matter}}\left[g\right]\right),\label{S_f}\\
S^{\mathrm{pot}} & = & \int\!\mathrm{d}t\mathrm{d}^{3}x\sqrt{-g}\,M_{g}^{2}m^{2}\sum_{n=0}^{4}c_{n}\,e_{n}\left(\bm{X}\right),\label{S_pot}\\
S^{\mathrm{com}} & = & \int\!\mathrm{d}^{4}x\sqrt{-\tilde{g}}\,P\big(\tilde{X},\phi\big),\label{S_com}
\end{eqnarray}
where $R[g]$ and $R[f]$ are Ricci scalar for $g_{\mu\nu}$ and $f_{\mu\nu}$, respectively. 
As in \cite{Gumrukcuoglu:2015nua}, in this work we consider the matter contents of $g_{\mu\nu}$ and $f_{\mu\nu}$ metrics to be  two cosmological constant: $\mathcal{L}^{\mathrm{matter}}[g] = -M_{g}^2\, \Lambda_g$ and $\mathcal{L}^{\mathrm{matter}}[f] = -M_{f}^2\, \Lambda_f$.
$S^{\mathrm{pot}}$ denotes the non-derivative potential interactions $S^{\mathrm{pot}}$ of the two metrics, where $\bm{X}$ stands for $X^{\mu}_{\phantom{\mu}\nu}$ and for a matrix $M_{\phantom{\mu}\nu}^{\mu}$, $e_{n}\left(\bm{M}\right)$ are the elementary symmetric polynomials defined by
			\begin{equation}
				e_{n}\left(\bm{M}\right)\equiv n!M_{[\mu_{1}}^{\mu_{1}}M_{\mu_{2}}^{\mu_{2}}\cdots M_{\mu_{n}]}^{\mu_{n}},
			\end{equation}
where the antisymmetrization is unnormalized. 
In (\ref{S_com}),  $\tilde{X}$ denoting the canonical kinetic term of $\phi$ in terms of the composite metric
	\begin{equation}
	\tilde{X}\equiv-\frac{1}{2}\tilde{g}^{\mu\nu}\partial_{\mu}\phi\partial_{\nu}\phi.\label{X_tilde}
	\end{equation}
In the following we will study this action on FLRW background and establish our parametrization for linear perturbations.

\section{Cosmological parametrization}

We parametrize the two metrics $g_{\mu\nu}$ and $f_{\mu\nu}$ to be
	\begin{eqnarray}
	g_{\mu\nu}\mathrm{d}x^{\mu}\mathrm{d}x^{\nu} & = & -N_{g}^{2}\left(e^{2A}-\left(e^{-\bm{H}}\right)^{ij}B_{i}B_{j}\right)\mathrm{d}t^{2}+2N_{g}a_{g}B_{i}\mathrm{d}t\mathrm{d}x^{i}+a_{g}^{2}\left(e^{\bm{H}}\right)_{ij}\mathrm{d}x^{i}\mathrm{d}x^{j},\label{metric_g}\\
	f_{\mu\nu}\mathrm{d}x^{\mu}\mathrm{d}x^{\nu} & = & -N_{f}^{2}\left(e^{2\varphi}-\left(e^{-\bm{\Gamma}}\right)^{ij}\Omega_{i}\Omega_{j}\right)\mathrm{d}t^{2}+2N_{f}a_{f}\Omega_{i}\mathrm{d}t\mathrm{d}x^{i}+a_{f}^{2}\left(e^{\bm{\Gamma}}\right)_{ij}\mathrm{d}x^{i}\mathrm{d}x^{j},\label{metric_f}
	\end{eqnarray}
where $N_{g}$, $a_{g}$, $N_f$ and $a_f$ are functions of time only, and the matrix exponentials are defined perturbatively as $\left(e^{\bm{H}}\right)_{ij}\equiv\delta_{ij}+H_{ij}+\frac{1}{2}H_{i}^{\phantom{i}k}H_{kj}+\mathcal{O}\left(H^{3}\right)$ and $\left(e^{-\bm{H}}\right)^{ij}=\delta^{ij}-H^{ij}+\frac{1}{2}H_{\phantom{i}k}^{i}H^{kj}+\mathcal{O}\left(H^{3}\right)$, etc.
Throughout this paper, spatial indices are raised and lowered by $\delta_{ij}$ and $\delta^{ij}$.
We further decompose (with $\partial^{2}\equiv\delta^{ij}\partial_{i}\partial_{j}$)
\begin{eqnarray}
B_{i} & \equiv & \partial_{i}B+S_{i},\label{B_dec}\\
H_{ij} & \equiv & 2\zeta\,\delta_{ij}+\left(\partial_{i}\partial_{j}-\frac{1}{3}\delta_{ij}\partial^{2}\right)E+\partial_{(i}F_{j)}+h_{ij},\label{H_dec}\\
\Omega_{i} & \equiv & \partial_{i}\omega+\sigma_{i},\label{Omega_dec}\\
\Gamma_{ij} & \equiv & 2\psi\,\delta_{ij}+\left(\partial_{i}\partial_{j}-\frac{1}{3}\delta_{ij}\partial^{2}\right)\chi+\partial_{(i}\xi_{j)}+\gamma_{ij},\label{Gamma_dec}
\end{eqnarray}
with $\partial_{(i}F_{j)}\equiv\frac{1}{2}\left(\partial_{i}F_{j}+\partial_{j}F_{i}\right)$,
etc, and
\begin{equation}
\partial^{i}S_{i}=\partial^{i}F_{i}=\partial^{i}\sigma_{i}=\partial^{i}\xi_{i}=0,\qquad h_{\phantom{i}i}^{i}=\gamma_{\phantom{i}i}^{i}=0,\qquad\partial^{i}h_{ij}=\partial^{i}\gamma_{ij}=0.
\end{equation}
Accordingly, it is convenient to parametrize the composite metric to be
	\begin{equation}
		\tilde{g}_{\mu\nu}\mathrm{d}x^{\mu}\mathrm{d}x^{\nu}=-\tilde{N}^{2}\left(e^{2\tilde{A}}-(e^{-\tilde{\bm{H}}})^{ij}\tilde{B}_{i}\tilde{B}_{j}\right)\mathrm{d}t^{2}+2\tilde{N}\tilde{a}\tilde{B}_{i}\mathrm{d}t\mathrm{d}x^{i}+a^{2}(e^{\tilde{\bm{H}}})_{ij}\mathrm{d}x^{i}\mathrm{d}x^{j},
	\end{equation}
where
	\begin{equation}
		\tilde{N}\equiv\alpha\,N+\beta\,N_{f},\qquad\tilde{a}\equiv\alpha\,a+\beta\,a_{f}.
	\end{equation}
Similar to (\ref{B_dec})-(\ref{Gamma_dec}), we may also decompose
	\begin{equation}
		\tilde{B}_{i}\equiv\partial_{i}\tilde{B}+\tilde{S}_{i},\qquad\tilde{H}_{ij}\equiv2\tilde{\zeta}\,\delta_{ij}+\left(\partial_{i}\partial_{j}-\frac{1}{3}\delta_{ij}\partial^{2}\right)\tilde{E}+\partial_{(i}\tilde{F}_{j)}+\tilde{h}_{ij},
	\end{equation}
with $\partial^i \tilde{S}_i = \partial^i \tilde{F}_i = \partial^i \tilde{h}_{ij}= \delta^{ij}\tilde{h}_{ij} =0$.
Note $\tilde{A}$ etc. are expressed in terms of $\{A,B_{i},H_{ij},\varphi,\Omega_{i},\Gamma_{ij}\}$ as
	\begin{equation}
		\tilde{A}=\sum_{n=1}\tilde{A}^{(n)}\left(A,B_{i},H_{ij},\varphi,\Omega_{i},\Gamma_{ij}\right)
	\end{equation}
etc., where $n$ denotes the order in $\{A,B_{i},H_{ij},\varphi,\Omega_{i},\Gamma_{ij}\}$.
At the linear order, we have, for the scalar modes,
	\begin{eqnarray}
	\tilde{A}^{(1)} & = & \alpha\frac{N}{\tilde{N}}A+\beta\frac{N_{f}}{\tilde{N}}\varphi,\label{At1}\\
	\tilde{B}^{(1)} & = & \alpha\,r_{1}B+\beta\,r_{2}\omega,\label{Bst1}\\
	\tilde{\zeta}^{(1)} & = & \alpha\frac{a}{\tilde{a}}\zeta+\beta\frac{a_{f}}{\tilde{a}}\psi,\label{zetat1}\\
	\tilde{E}^{(1)} & = & \alpha\frac{a}{\tilde{a}}E+\beta\frac{a_{f}}{\tilde{a}}\chi,\label{Et1}
	\end{eqnarray}
with
	\begin{equation}
	r_{1}\equiv\frac{aN\left(N_{f}\tilde{a}+a_{f}\tilde{N}\right)}{\left(Na_{f}+aN_{f}\right)\tilde{a}\tilde{N}},\qquad r_{2}\equiv\frac{a_{f}N_{f}\left(N\tilde{a}+a\tilde{N}\right)}{\left(Na_{f}+aN_{f}\right)\tilde{a}\tilde{N}},\label{r1r2}
	\end{equation}
	for the vector modes,
	\begin{equation}
	\tilde{S}_{i}^{(1)}=\alpha\,r_{1}S_{i}+\beta\,r_{2}\sigma_{i},\qquad\tilde{F}_{i}^{(1)}=\alpha\frac{a}{\tilde{a}}F_{i}+\beta\frac{a_{f}}{\tilde{a}}\xi_{i},\label{SiFit1}
	\end{equation}
	and for the tensor modes
	\begin{equation}
	\tilde{h}_{ij}^{(1)}=\alpha\frac{a}{\tilde{a}}h_{ij}+\beta\frac{a_{f}}{\tilde{a}}\gamma_{ij}.\label{hijt1}
	\end{equation}

The background equations of motion can be determined by requiring the vanishing of the first order action of $A$, $\zeta$, $\varphi$, $\psi$ and $\delta\phi$, which is given by
	\begin{equation}
		S_{1}=\int\!\mathrm{d}t\mathrm{d}^{3}x\,N_{g}a_{g}^{3}\left(\mathcal{E}_{A}A+\mathcal{E}_{\zeta}\,3\zeta+\mathcal{E}_{\varphi}\varphi+\mathcal{E}_{\psi}3\psi+\frac{\tilde{N}\tilde{a}^{3}}{N_{g}a_{g}^{3}}\mathcal{E}_{\phi}\,\delta\phi\right).
	\end{equation}
The set of equations of motion are thus given by
	\begin{eqnarray}
	\mathcal{E}_{A} & \equiv & M_{g}^{2}\left(3H_{g}^{2}-\Lambda_{g}\right)+\mathcal{E}_{A}^{\mathrm{pot}}+\alpha\frac{\tilde{a}^{3}}{a_{g}^{3}}\left(P-2\tilde{X}P_{,\tilde{X}}\right)=0,\label{bgeom_A}\\
	\mathcal{E}_{\zeta} & \equiv & M_{g}^{2}\left(3H_{g}^{2}+\frac{2}{N_{g}}\frac{\mathrm{d}H_{g}}{\mathrm{d}t}-\Lambda_{g}\right)+\mathcal{E}_{\zeta}^{\mathrm{pot}}+\alpha\frac{\tilde{N}\tilde{a}^{2}}{N_{g}a_{g}^{2}} P =0,\label{bgeom_zeta}\\
	\mathcal{E}_{\varphi} & \equiv & \frac{N_{f}a_{f}^{3}}{N_{g}a_{g}^{3}}M_{f}^{2}\left(3H_{f}^{2}-\Lambda_{f}\right)+\mathcal{E}_{\varphi}^{\mathrm{pot}}+\beta\frac{N_{f}\tilde{a}^{3}}{N_{g}a_{g}^{3}}
	\left(P-2\tilde{X}P_{,\tilde{X}}\right)=0,\label{bgeom_vphi}\\
	\mathcal{E}_{\psi} & \equiv & \frac{N_{f}a_{f}^{3}}{N_{g}a_{g}^{3}}M_{f}^{2}\left(3H_{f}^{2}+\frac{2}{N_{f}}\frac{\mathrm{d}H_{f}}{\mathrm{d}t}-\Lambda_{f}\right)+\mathcal{E}_{\psi}^{\mathrm{pot}}+\beta\frac{\tilde{N}\tilde{a}^{2}a_{f}}{N_{g}a_{g}^{3}}
	P=0,\label{bgeom_psi}
	\end{eqnarray}
where $P_{,\tilde{X}}$ is the shorthand for $\partial P/\partial \tilde{X}$, the $H_{g}$ and $H_{f}$ are the Hubble parameter associated with the two metrics respectively, i.e.,
	\begin{equation}
		H_{g}\equiv\frac{1}{N_{g}a_{g}}\frac{\mathrm{d}a_{g}}{\mathrm{d}t}, \qquad H_{f}\equiv\frac{1}{N_{f}a_{f}}\frac{\mathrm{d}a_{f}}{\mathrm{d}t}.
	\end{equation}
In the above,
	\begin{eqnarray}
	\mathcal{E}_{A}^{\mathrm{pot}} & = & M_{g}^{2}m^{2}\left(c_{0}+3\frac{a_{f}}{a_{g}}c_{1}+6\frac{a_{f}^{2}}{a_{g}^{2}}c_{2}+6\frac{a_{f}^{3}}{a_{g}^{3}}c_{3}\right),\label{Epot_A}\\
	\mathcal{E}_{\zeta}^{\mathrm{pot}} & = & b_{1}+\frac{a_{g}N_{f}}{N_{g}a_{f}}b_{2},\label{Epot_zeta}\\
	\mathcal{E}_{\varphi}^{\mathrm{pot}} & = & M_{g}^{2}m^{2}\frac{N_{f}}{N_{g}}\left(c_{1}+6\frac{a_{f}}{a_{g}}c_{2}+18\frac{a_{f}^{2}}{a_{g}^{2}}c_{3}+24\frac{a_{f}^{3}}{a_{g}^{3}}c_{4}\right),\label{Epot_vphi}\\
	\mathcal{E}_{\psi}^{\mathrm{pot}} & = & b_{2}+b_{3},\label{Epot_psi}
	\end{eqnarray}
where we have introduced
	\begin{eqnarray}
	b_{1} & \equiv & M_{g}^{2}m^{2}\left(c_{0}+2\frac{a_{f}}{a_{g}}c_{1}+2\frac{a_{f}^{2}}{a_{g}^{2}}c_{2}\right),\label{b1}\\
	b_{2} & \equiv & M_{g}^{2}m^{2}\frac{a_{f}}{a_{g}}\left(c_{1}+4\frac{a_{f}}{a_{g}}c_{2}+6\frac{a_{f}^{2}}{a_{g}^{2}}c_{3}\right),\label{b2}\\
	b_{3} & \equiv & 2M_{g}^{2}m^{2}\frac{N_{f}a_{f}}{N_{g}a_{g}}\left(c_{2}+6\frac{a_{f}}{a_{g}}c_{3}+12\frac{a_{f}^{2}}{a_{g}^{2}}c_{4}\right),\label{b3}
	\end{eqnarray}
for later convenience.
The equation of motion for the scalar field is given by
	\begin{equation}
	\mathcal{E}_{\phi}\equiv P_{,\phi}-\frac{1}{\tilde{N}\tilde{a}^{3}}\frac{\mathrm{d}}{\mathrm{d}t}\left(\frac{\tilde{a}^{3}}{\tilde{N}}\frac{\mathrm{d}\bar{\phi}}{\mathrm{d}t}P_{,\tilde{X}}\right),\label{Ecom_phi}
	\end{equation}

\section{Cosmological perturbations}

The quadratic action for the two tensor perturbations $h_{ij}$ and $\gamma_{ij}$ is given by
	\begin{eqnarray}
	S_{2}^{\mathrm{tensor}} & = & \frac{1}{8}\int\!\mathrm{d}t\frac{\mathrm{d}^{3}k}{\left(2\pi\right)^{3}}\bigg[ N_{g}a_{g}^{3}M_{g}^{2}\left(\frac{1}{N_{g}^{2}}\dot{h}_{ij}^{2}-\frac{k^{2}}{a^{2}}h_{ij}^{2}\right)+N_{f}a_{f}^{3}M_{f}^{2}\left(\frac{1}{N_{f}^{2}}\dot{\gamma}_{ij}^{2}-\frac{k^{2}}{a_{f}^{2}}\gamma_{ij}^{2}\right)\nonumber \\
	&  & \hspace{4em}+N_{g}a_{g}^{3}\mathcal{M}^{2}\left(h_{ij}-\gamma_{ij}\right)\left(h^{ij}-\gamma^{ij}\right)\bigg].\label{S2_ten}
	\end{eqnarray}
where a dot denotes derivative with respect to $t$,
	\begin{equation}
	\mathcal{M}^{2}\equiv\frac{a_{f}}{a_{g}}\left[M_{g}^{2}m^{2}\left(c_{1}+2\frac{a_{f}}{a_{g}}c_{2}+2\frac{N_{f}}{N_{g}}\left(c_{2}+3\frac{a_{f}}{a_{g}}c_{3}\right)\right)+\alpha\beta\frac{\tilde{N}\tilde{a}}{N_{g}a_{g}}P\right].\label{Mcal}
	\end{equation}

The quadratic action for the four vector modes $S_i$, $F_i$, $\sigma_i$ and $\xi_i$ is given by
	\begin{eqnarray}
	S_{2}^{\mathrm{vector}} & = & \int\!\mathrm{d}t\frac{\mathrm{d}^{3}k}{\left(2\pi\right)^{3}}\Bigg\{\frac{1}{4}N_{g}a_{g}^{3}\,M_{g}^{2}k^{2}\left(\frac{1}{a_{g}}S_{i}-\frac{1}{2N_{g}}\dot{F}_{i}\right)^{2}+\frac{1}{4}N_{f}a_{f}^{3}\,M_{f}^{2}k^{2}\left(\frac{1}{a_{f}}\sigma_{i}-\frac{1}{2N_{f}}\dot{\xi}_{i}\right)^{2}\nonumber \\
	&  & \hspace{4em}-\frac{1}{2}N_{g}a_{g}^{3}\mathcal{C}\left(S_{i}-\frac{a_{g}N_{f}}{N_{g}a_{f}}\sigma_{i}\right)^{2}+\frac{N_{g}a_{g}^{3}}{16}\mathcal{M}^{2}k^{2}\left(F_{i}-\xi_{i}\right)^{2}\Bigg\},\label{S2_vec}
	\end{eqnarray}
where  $\mathcal{M}^2$ is given in (\ref{Mcal}) and we also introduce
	\begin{equation}
	\mathcal{C}\equiv\frac{1}{1+\frac{a_{g}N_{f}}{N_{g}a_{f}}}b_{2}+\frac{\alpha\beta}{\left(1+\frac{a_{g}N_{f}}{N_{g}a_{f}}\right)^{2}}\frac{\tilde{N}\tilde{a}a_{f}}{N_{g}a_{g}^{2}}\left[\left(1+\frac{\tilde{a}N_{f}}{\tilde{N}a_{f}}+\frac{N_{g}\tilde{a}}{\tilde{N}a_{g}}\right)\left(P-2\tilde{X}P_{,\tilde{X}}\right)-P\right], \label{Ccal}
	\end{equation}
with $b_2$ given in (\ref{b2}) for short. Since the vector modes $S_i$ and $\sigma_i$ have no dynamics in (\ref{S2_vec}), we may solve them in terms of $F_i$ and $\xi_i$ and arrive at the reduced action for $F_i$ and $\xi_i$, which is given by
	\begin{equation}
	S_{2}^{\mathrm{vector}}=\frac{1}{16}\int\!\mathrm{d}t\frac{\mathrm{d}^{3}k}{\left(2\pi\right)^{3}}\,N_{g}a_{g}^{3}\,k^{2}\left[\mathcal{G}_{\mathrm{v}}\frac{1}{N_{g}^{2}}\left(\partial_{t}\left(F_i-\xi_i\right)\right)^{2}+\mathcal{M}^{2}\left(F_i-\xi_i\right)^{2}\right],\label{S2_vec_f}
	\end{equation}
with
	\begin{equation}
		\mathcal{G}_{\mathrm{v}}=\left(\frac{a_{g}^{3}N_{f}}{a_{f}^{3}N_{g}}\frac{1}{M_{f}^{2}}+\frac{1}{M_{g}^{2}}-\frac{1}{2\mathcal{C}}\frac{k^{2}}{a_{g}^{2}}\right)^{-1}.
	\end{equation}
From (\ref{S2_vec_f}) it is transparent that there are two vectorial degrees of freedom giving that $\beta\neq 0$, which can be identified as $F_i - \xi_i$. For the stability condition we have to impose $\mathcal{G}_{\mathrm{v}}>0$.

We study now the linear stability of the scalar modes in our model.
Initially we have 9 scalar modes, of which four ($A$, $B$, $\zeta$ and $E$) are from $g_{\mu\nu}$, four ($\varphi$, $\omega$, $\psi$ and $\chi$) are from $f_{\mu\nu}$, and one is the perturbation of the scalar field $\delta\phi$.
In order to simplify the calculation, we choose a gauge in which $\delta\phi = \chi =0$.
In the residual 7 modes, only 2 modes are dynamical, which can be conveniently chosen to be
	\begin{equation}
	\left(\begin{array}{c}
	V_{1}\\
	V_{2}
	\end{array}\right)\equiv\left(\begin{array}{c}
	Q\\
	E
	\end{array}\right).
	\end{equation}
with
	\begin{equation}
		Q=\zeta+\frac{k^{2}}{6}E+\frac{\beta H_{g}}{\alpha H_{f}}\psi. \label{Q_def}
	\end{equation}
After some manipulations, the final quadratic action for these two scalar modes takes the following general structure (in matrix form)
	\begin{eqnarray}
	S_{2}^{\mathrm{scalar}}=\frac{1}{2}\int\!\mathrm{d}t\frac{\mathrm{d}^{3}k}{\left(2\pi\right)^{3}}\left(\dot{V}^{T}\bm{\mathcal{G}}\dot{V}+\dot{V}^{T}\bm{\mathcal{F}}V+V^{T}\bm{\mathcal{W}}V\right),\label{S2_sca}
	\end{eqnarray}
where $\mathcal{G}_{mn}$ and $\mathcal{W}_{mn}$ are symmetric while $\mathcal{F}_{mn}$
is antisymmetric, which are given by
	\begin{eqnarray}
	\mathcal{G}_{mn} & = & \Xi_{mn}-\frac{1}{\mathcal{D}}\mathcal{A}_{m}\mathcal{A}_{n}, \label{Gcal}\\
	\mathcal{F}_{12} \equiv -\mathcal{F}_{21} & = & \mathcal{A} - \frac{1}{\mathcal{D}}\left(\mathcal{D}_{1}\mathcal{A}_{2}-\mathcal{D}_{2}\mathcal{A}_{1}\right), \label{Fcal}\\
	\mathcal{W}_{mn} & = & \mathcal{B}_{mn}-\frac{1}{\mathcal{D}}\mathcal{D}_{m}\mathcal{D}_{n}-\frac{1}{2}\frac{\mathrm{d}}{\mathrm{d}t}\left[\frac{1}{\mathcal{D}}\left(\mathcal{D}_{m}\mathcal{A}_{n}+\mathcal{D}_{n}\mathcal{A}_{m}\right)\right], \label{Wcal}
	\end{eqnarray}
with $ m,n=1,2$. In (\ref{Gcal})-(\ref{Fcal}),
we have
\begin{eqnarray}
\mathcal{D} & = & \frac{\beta^{2}}{\alpha^{2}}\frac{H_{g}^{2}}{H_{f}^{2}}\left[\left(\frac{\mathrm{d}}{\mathrm{d}t}\left(\ln\frac{H_{g}}{H_{f}}\right)\right)^{2}\Xi_{11}+\Xi_{44}-\frac{\mathrm{d}\Xi_{14}}{\mathrm{d}t}\right]-\frac{\mathrm{d}\Xi_{36}}{\mathrm{d}t}+\Xi_{66}\\
&  & +\frac{\beta}{\alpha}\frac{H_{g}}{H_{f}}\left[-\frac{\mathrm{d}}{\mathrm{d}t}\left(\ln\frac{H_{g}}{H_{f}}\right)\left(\Xi_{16}-\Xi_{34}\right)+\frac{\mathrm{d}\Xi_{16}}{\mathrm{d}t}-2\Xi_{46}+\frac{\mathrm{d}\Xi_{34}}{\mathrm{d}t}\right],
\label{Dcal}
\end{eqnarray}
\begin{eqnarray}
\mathcal{D}_{1} & \equiv & \Xi_{46}-\frac{\mathrm{d}\Xi_{34}}{\mathrm{d}t}+\frac{\beta H_{g}}{\alpha H_{f}}\left(\frac{\mathrm{d}\Xi_{14}}{\mathrm{d}t}-\Xi_{44}\right),\label{Dcal1}\\
\mathcal{D}_{2} & \equiv & -\frac{\mathrm{d}\Xi_{35}}{\mathrm{d}t}+\Xi_{56}+\frac{k^{2}}{6}\Big(\frac{\mathrm{d}\Xi_{34}}{\mathrm{d}t}-\Xi_{46}\Big)+\frac{\beta H_{g}}{\alpha H_{f}}\left[\frac{k^{2}}{6}\Big(\Xi_{44}-\frac{\mathrm{d}\Xi_{14}}{\mathrm{d}t}\Big)-\Xi_{45}+\frac{\mathrm{d}\Xi_{15}}{\mathrm{d}t}\right],\label{Dcal2}
\end{eqnarray}
and
\begin{eqnarray}
\mathcal{A}_{1} & \equiv & \Xi_{34}-\Xi_{16}+\frac{\beta}{\alpha}\frac{\mathrm{d}}{\mathrm{d}t}\left(\frac{H_{g}}{H_{f}}\right)\Xi_{11},\label{Acal1}\\
\mathcal{A}_{2} & \equiv & \Xi_{35}-\Xi_{26}+\frac{k^{2}}{6}\left(\Xi_{16}-\Xi_{34}\right)+\frac{\beta H_{g}}{\alpha H_{f}}\left[\Xi_{24}-\Xi_{15}+\frac{\mathrm{d}}{\mathrm{d}t}\Big(\ln\frac{H_{g}}{H_{f}}\Big)\Big(\Xi_{12}-\frac{k^{2}}{6}\Xi_{11}\Big)\right],\label{Acal2}\\
\mathcal{A} & \equiv & \Xi_{15}-\Xi_{24},\label{Acal}
\end{eqnarray}
and
\begin{eqnarray}
\mathcal{B}_{11} & \equiv & \Xi_{44}-\frac{\mathrm{d}\Xi_{14}}{\mathrm{d}t},\label{Bcal11}\\
\mathcal{B}_{12}\equiv\mathcal{B}_{21} & \equiv & \Xi_{45}-\frac{k^{2}}{6}\Xi_{44}-\frac{1}{2}\frac{\mathrm{d}}{\mathrm{d}t}\left(\Xi_{15}+\Xi_{24}-\frac{k^{2}}{3}\Xi_{14}\right),\label{Bcal12}\\
\mathcal{B}_{22} & \equiv & \frac{k^{4}}{36}\Xi_{44}-\frac{k^{2}}{3}\Xi_{45}+\Xi_{55}-\frac{\mathrm{d}}{\mathrm{d}t}\left(\frac{k^{4}}{36}\Xi_{14}-\frac{k^{2}}{6}\Xi_{15}-\frac{k^{2}}{6}\Xi_{24}+\Xi_{25}\right),\label{Bcal22}
\end{eqnarray}
where $\Xi_{ij}$ with $i,j=1,\cdots,6$ are given in Appendix \ref{app:sca}. Up to now, no approximation is made in deriving the above expressions.

Unlike the tensor and vector modes, the lengthy expressions in the above make the analysis for the scalar modes rather cumbersome.
In the following, we analyze the instabilities in the small scale limit $k\rightarrow \infty$.
For the kinetic terms, we have
	\begin{equation}
		\mathcal{G}_{11}=\hat{\mathcal{G}}_{11}+\mathcal{O}\left(k^{-2}\right),\qquad\text{and}\qquad\mathcal{G}_{22}=k^{2}\hat{\mathcal{G}}_{22}+\mathcal{O}\left(k^{0}\right),
	\end{equation}
where
	\begin{equation}
		\hat{\mathcal{G}}_{11}=\frac{\alpha^{2}\left(\frac{\mathrm{d}\bar{\Phi}}{\mathrm{d}t}\right)^{2}\tilde{a}^{3}}{\tilde{N}^{3}H_{g}^{2}}\left(P_{,\tilde{X}}+2\tilde{X}P_{,\tilde{X}\tilde{X}}\right),
	\end{equation}
and
\begin{equation}
	\hat{\mathcal{G}}_{22}= -\frac{\mathcal{C}a_{g}^{5}}{4N_{g}}-\frac{1}{\hat{\mathcal{D}}}\left(\hat{\mathcal{A}}_{2}\right)^{2},
\end{equation}
with
\begin{eqnarray}
\hat{\mathcal{D}} & = & M_{g}^{2}N_{g}a_{g}\Bigg[\frac{2\beta^{2}}{\alpha^{2}}\left(\frac{1}{H_{f}}\frac{2}{N_{g}}\frac{\mathrm{d}}{\mathrm{d}t}\left(\frac{H_{g}}{H_{f}}\right)+\frac{H_{g}^{2}}{H_{f}^{2}}\right)+2\frac{a_{f}}{a_{g}}\frac{M_{f}^{2}}{M_{g}^{2}}\frac{N_{f}}{N_{g}}-\frac{\mathcal{C}a_{g}^{4}N_{f}^{2}}{a_{f}^{4}H_{f}^{2}N_{g}^{2}M_{g}^{2}}\left(1+\frac{\beta a_{f}^{2}N_{g}}{\alpha a_{g}^{2}N_{f}}\right)^{2}\nonumber \\
&  & \qquad-\frac{1}{M_{g}^{2}a_{g}}\frac{2}{N_{g}}\frac{\mathrm{d}}{\mathrm{d}t}\left(\frac{a_{f}}{H_{f}}M_{f}^{2}+\frac{\beta^{2}}{\alpha^{2}}\frac{a_{g}H_{g}}{H_{f}^{2}}M_{g}^{2}\right)\Bigg],\label{Dcal_hat}
\end{eqnarray}
and
	\begin{equation}
	\hat{\mathcal{A}}_{2}=\frac{a_{g}^{5}N_{f}}{2\alpha a_{f}^{2}H_{f}N_{g}}\left[\mathcal{C}\left(\beta\frac{a_{f}^{2}N_{g}}{a_{g}^{2}N_{f}}+\alpha\right)+\beta\frac{a_{f}^{2}N_{g}}{a_{g}^{2}N_{f}}\left(b_{1}-M_{g}^{2}\Lambda_{g}+3M_{g}^{2}H_{g}^{2}\right)-\alpha b_{2}\frac{a_{f}N_{g}}{a_{g}N_{f}}\right].\label{Acal_2_hat}
	\end{equation}
It can also be verified that $\mathcal{G}_{12} \sim \mathcal{O}(k^0)$.
Thus in the large $k$ limit, the no ghost condition concerning the kinetic terms requires that $P_{,\tilde{X}}+2\tilde{X}P_{,\tilde{X}\tilde{X}}>0$ as well as
	\begin{equation}
		\frac{\mathcal{C}a_{g}^{5}}{4N_{g}}+\frac{1}{\hat{\mathcal{D}}}\left(\hat{\mathcal{A}}_{2}\right)^{2}<0.
	\end{equation}
These results can be compared with those derived in \cite{Gumrukcuoglu:2015nua}.

As for the gradient terms, in the large $k$ limit we have
	\begin{equation}
		\mathcal{W}_{11}=k^{2}\hat{\mathcal{W}}_{11}+\mathcal{O}\left(k^{0}\right),\qquad\mathcal{W}_{22}=k^{4}\hat{\mathcal{W}}_{22}+\mathcal{O}\left(k^{2}\right),
	\end{equation}
and $\mathcal{W}_{12}\sim \mathcal{O}(k^2)$, where
	\begin{equation}
		\hat{\mathcal{W}}_{11}=\frac{a_{g}N_{g}}{H_{g}^{2}}\left(2M_{g}^{2}\frac{1}{N_{g}}\frac{\mathrm{d}H_{g}}{\mathrm{d}t}-\mathcal{C}\right)-\frac{1}{\hat{\mathcal{D}}}\hat{\mathcal{D}}_{1}\hat{\mathcal{D}}_{2}, \label{Wcal11_hat}
	\end{equation}
and
	\begin{eqnarray}
	\hat{\mathcal{W}}_{22} & = & \frac{1}{4}a_{g}^{3}N_{g}\bigg[m^{2}M_{g}^{2}\left(c_{0}+c_{1}\left(\frac{a_{f}}{a_{g}}+\frac{N_{f}}{N_{g}}\right)+2c_{2}\frac{a_{f}}{a_{g}}\frac{N_{f}}{N_{g}}\right)\nonumber \\
	&  & +\alpha^{2}\frac{\tilde{a}\tilde{N}}{a_{g}N_{g}}P-M_{g}^{2}\Lambda_{g}+M_{g}^{2}\left(3H_{g}^{2}+\frac{2}{N_{g}}\frac{\mathrm{d}H_{g}}{\mathrm{d}t}\right)+\frac{1}{3}\mathcal{M}^{2}\bigg].
	\end{eqnarray}
In (\ref{Wcal11_hat}), $\hat{\mathcal{D}}$ is given in (\ref{Dcal_hat}), and
	\begin{equation}
		\hat{\mathcal{D}}_1 =\frac{a_{g}}{M_{g}^{2}H_{f}}\left[\frac{\mathcal{C}a_{g}^{2}N_{f}}{a_{f}^{2}H_{g}M_{g}^{2}}\left(1+\frac{\beta a_{f}^{2}N_{g}}{\alpha a_{g}^{2}N_{f}}\right)-2\frac{\beta}{\alpha}\frac{\mathrm{d}\ln H_{g}}{\mathrm{d}t}\right],
	\end{equation}	
and
	\begin{eqnarray}
	\hat{\mathcal{D}}_{2} & = & a_{g}^{3}N_{g}\Bigg\{\frac{\beta}{\alpha}\left[\frac{\alpha^{2}\tilde{a}^{2}}{a_{g}^{2}}\left(P-2\tilde{X}P_{,\tilde{X}}\right)+b_{1}+\mathcal{E}_{A}^{g}\right]\frac{1}{2H_{g}}\frac{1}{N_{g}}\frac{\mathrm{d}}{\mathrm{d}t}\left(\frac{H_{g}}{H_{f}}\right)+\frac{1}{2}\left(1+\frac{\beta H_{g}}{\alpha H_{f}}\right)\mathcal{M}^{2}\nonumber \\
	&  & +\frac{3\beta H_{g}}{2\alpha H_{f}}M_{g}^{2}\left(3H_{g}^{2}+\frac{2}{N_{g}}\frac{\mathrm{d}H_{g}}{\mathrm{d}t}-\Lambda_{g}\right)+\frac{3\alpha\beta\tilde{a}\tilde{N}P}{2a_{g}N_{g}}\left(\frac{H_{g}}{H_{f}}-\frac{a_{f}}{a_{g}}\right)\nonumber \\
	&  & +\frac{3}{2}m^{2}M_{g}^{2}\bigg[\frac{\beta H_{g}}{\alpha H_{f}}\left(c_{0}+c_{1}\Big(\frac{a_{f}}{a_{g}}+\frac{N_{f}}{N_{g}}\Big)+2c_{2}\frac{a_{f}}{a_{g}}\frac{N_{f}}{N_{g}}\right)\nonumber \\
	&  & \qquad-\frac{a_{f}}{a_{g}}\left(c_{1}+2c_{2}\Big(\frac{a_{f}}{a_{g}}+\frac{N_{f}}{N_{g}}\Big)+6c_{3}\frac{a_{f}}{a_{g}}\frac{N_{f}}{N_{g}}\right)\bigg]\nonumber \\
	&  & +\frac{\mathcal{C}}{4}\frac{1}{H_{f}H_{g}M_{g}^{2}}\left(\beta+\alpha\frac{N_{f}}{N_{g}}\frac{a_{g}^{2}}{a_{f}^{2}}\right)\bigg[-\alpha\frac{\tilde{a}^{2}}{a_{g}^{2}}\left(P-2\tilde{X}P_{,\tilde{X}}\right)\nonumber \\
	&  & \qquad-\frac{1}{\alpha}\left(b_{1}-M_{g}^{2}\Lambda_{g}+3M_{g}^{2}H_{g}^{2}\right)+\frac{a_{g}^{3}N_{f}H_{g}M_{g}^{2}}{a_{f}^{3}N_{g}H_{f}M_{f}^{2}}\left(\beta\frac{\tilde{a}^{2}}{a_{g}^{2}}\left(P-2\tilde{X}P_{,\tilde{X}}\right)+\frac{a_{g}b_{2}}{\alpha a_{f}}\right)\bigg]\Bigg\}\nonumber \\
	&  & -\frac{1}{2}\frac{\mathrm{d}}{\mathrm{d}t}\left[a_{g}^{3}\left(\frac{\beta}{\alpha H_{f}}\left(b_{1}-M_{g}^{2}\Lambda_{g}+3M_{g}^{2}H_{g}^{2}\right)-\frac{a_{g}b_{2}}{a_{f}H_{f}}\right)\right].
	\end{eqnarray}
Thus, in the large $k$ limit, the absence of gradient instability requires
	\begin{equation}
		\hat{\mathcal{W}}_{11} >0, \qquad \text{and} \qquad  \hat{\mathcal{W}}_{22} >0. \label{Wcalpos}
	\end{equation}
The propagating speeds of the two scalar modes are given by the eigenvalues of $\bm{\mathcal{G}}^{-1}\bm{\mathcal{W}}$, which correspond to
	\begin{equation}
		c_{1}^{2}=\frac{\hat{\mathcal{W}}_{11}}{\hat{\mathcal{G}}_{11}}\qquad\text{and}\qquad c_{2}^{2}=\frac{\hat{\mathcal{W}}_{22}}{\hat{\mathcal{G}}_{22}}
	\end{equation}
in the same limit.

\section{Conclusion}

In this work, we investigate the cosmological perturbation analysis of the bimetric theory with a scalar field coupled simultaneously to both metrics in terms of a composite metric. The scalar field represents the matter field that lives on both metrics. 

The ghost and gradient instabilities of the tensor and vector modes as well as the ghost instabilities of the scalar modes of the same model have been analyzed in \cite{Gumrukcuoglu:2015nua} for some concrete background evolution, while in this work we complete the analysis by presenting the full quadratic action for the scalar modes (\ref{S2_sca}) as well as the conditions for the absence of gradient instabilities as in (\ref{Wcalpos}) on general background evolution in the presence of matter fields.
Although in this work we focus on the small scale limit $k\rightarrow 0$ due to the lengthy expressions, the results presented in this work enable one to make further analysis in different limits as well as upon concrete background solutions.

Moreover, we consider only the coupling of the scalar field to the composite metric in a minimal way, while in principle one may consider non-minimal derivative couplings as was pointed in \cite{Gao:2016vtt}. This bimetric model with doubly coupled matter fields offers an interesting cosmological framework. In one branch of solutions, in which the Hubble rates are proportional to each other, this interesting phenomenology is plagued by the ghost and gradient instabilities as was shown in \cite{Gumrukcuoglu:2015nua}. However, in the other branch of background cosmology with the algebraical ratio between the scale factors of the two metrics there are no ghost instabilities associated with the vector and scalar perturbations. Here, we also show the conditions for the absence of the gradient instabilities for the scalar perturbations, which were lacking in the literature. Fulfilling all these instability conditions, this branch of solutions still offers promising dark energy model, which has a very rich phenomenology \cite{Brax:2016ssf}.


\acknowledgments
We would like to thank S. Mukohyama for useful discussions.
X.G. was supported by JSPS Grant-in-Aid for Scientific Research No. 15H02082 and partly by JSPS Grant-in-Aid for Scientific Research No. 25287054 and 26610062.
LH acknowledges financial support from Dr. Max R\"ossler, the Walter Haefner Foundation and the ETH Zurich Foundation.

\appendix

\section{Expressions of $\Xi_{ab}$} \label{app:sca}

The expression of $\Xi_{ab}$ with $a,b=1,\cdots,6$ are given by
\begin{equation}
\Xi_{11}=-\frac{1}{\Delta}\frac{16}{\tilde{N}^{3}}\alpha^{2}\left(\frac{\mathrm{d}\bar{\phi}}{\mathrm{d}t}\right)^{2}k^{6}\tilde{a}^{3}a_{f}a_{g}^{4}g_{\phi\phi}H_{f}^{2}M_{f}^{2}M_{g}^{2}N_{f}^{2}N_{g}\left[a_{f}^{3}M_{f}^{2}N_{g}\left(3\mathcal{C}a_{g}^{2}-2k^{2}M_{g}^{2}\right)+3\mathcal{C}a_{g}^{5}M_{g}^{2}N_{f}\right]
\end{equation}
\begin{equation}
\Xi_{12}=\frac{1}{\Delta}\frac{8}{3\tilde{N}^{3}}\alpha\left(\frac{\mathrm{d}\bar{\phi}}{\mathrm{d}t}\right)^{2}k^{8}\tilde{a}^{3}a_{f}a_{g}^{4}g_{\phi\phi}H_{f}M_{f}^{2}M_{g}^{4}N_{f}^{2}N_{g}\left[2\alpha k^{2}a_{f}^{3}H_{f}M_{f}^{2}N_{g}-3\mathcal{C}a_{g}^{5}N_{f}(\alpha H_{f}+\beta H_{g})\right],
\end{equation}
\begin{eqnarray}
\Xi_{14} & = & \frac{1}{\Delta}\frac{8k^{6}a_{g}^{4}M_{g}^{2}N_{f}^{2}}{\tilde{N}^{3}a_{f}^{2}}\bigg\{\mathcal{C}\left(\frac{\mathrm{d}\bar{\phi}}{\mathrm{d}t}\right)^{2}\tilde{a}^{3}a_{g}^{2}g_{\phi\phi}(\alpha F_{24}N_{g}-\beta F_{14}N_{f})\left(\alpha a_{f}^{3}H_{f}M_{f}^{2}N_{g}-\beta a_{g}^{3}H_{g}M_{g}^{2}N_{f}\right)\nonumber \\
&  & +F_{14}\tilde{N}^{3}a_{f}^{3}H_{f}^{2}H_{g}M_{f}^{2}\left(a_{f}^{3}M_{f}^{2}N_{g}\left(2k^{2}M_{g}^{2}-3\mathcal{C}a_{g}^{2}\right)-3\mathcal{C}a_{g}^{5}M_{g}^{2}N_{f}\right)\bigg\},
\end{eqnarray}
\begin{eqnarray}
\Xi_{15} & = & \frac{1}{\Delta}\frac{8k^{10}a_{g}^{5}M_{g}^{4}N_{f}^{2}N_{g}}{3\tilde{N}^{3}a_{f}^{2}}\bigg\{\beta\mathcal{C}\left(\frac{\mathrm{d}\bar{\phi}}{\mathrm{d}t}\right)^{2}\tilde{a}^{3}a_{g}^{2}g_{\phi\phi}N_{f}\left(\beta a_{g}^{3}H_{g}M_{g}^{2}N_{f}-\alpha a_{f}^{3}H_{f}M_{f}^{2}N_{g}\right)\nonumber \\
&  & +\tilde{N}^{3}a_{f}^{3}H_{f}^{2}H_{g}M_{f}^{2}\left[a_{f}^{3}M_{f}^{2}N_{g}\left(2k^{2}M_{g}^{2}-3\mathcal{C}a_{g}^{2}\right)-3\mathcal{C}a_{g}^{5}M_{g}^{2}N_{f}\right]\bigg\},
\end{eqnarray}
\begin{eqnarray}
\Xi_{16} & = & \frac{1}{\Delta}\frac{8k^{6}a_{g}^{4}M_{g}^{2}N_{f}{}^{2}}{\tilde{N}^{3}a_{f}^{2}}\bigg\{\mathcal{C}\left(\frac{\mathrm{d}\bar{\phi}}{\mathrm{d}t}\right)^{2}\tilde{a}^{3}a_{g}^{2}g_{\phi\phi}\left(\alpha F_{26}N_{g}-\beta F_{16}N_{f}\right)\left(\alpha a_{f}^{3}H_{f}M_{f}^{2}N_{g}-\beta a_{g}^{3}H_{g}M_{g}^{2}N_{f}\right)\nonumber \\
&  & +F_{16}\tilde{N}^{3}a_{f}^{3}H_{f}^{2}H_{g}M_{f}^{2}\left[a_{f}^{3}M_{f}^{2}N_{g}\left(2k^{2}M_{g}^{2}-3\mathcal{C}a_{g}^{2}\right)-3\mathcal{C}a_{g}^{5}M_{g}^{2}N_{f}\right]\bigg\},
\end{eqnarray}
\begin{eqnarray}
\Xi_{22} & = & -\frac{1}{\Delta}\frac{4}{9\tilde{N}^{3}}k^{10}a_{f}a_{g}^{4}H_{f}M_{f}^{2}M_{g}^{2}N_{f}^{2}N_{g}\bigg\{9\mathcal{C}\tilde{N}^{3}a_{f}^{3}a_{g}^{5}H_{f}H_{g}^{2}M_{f}^{2}M_{g}^{2}\nonumber \\
&  & -\alpha\left(\frac{\mathrm{d}\bar{\phi}}{\mathrm{d}t}\right)^{2}\tilde{a}^{3}g_{\phi\phi}\left[\alpha a_{f}^{3}H_{f}M_{f}^{2}N_{g}\left(2k^{2}M_{g}^{2}+3\mathcal{C}a_{g}^{2}\right)-3\mathcal{C}a_{g}^{5}M_{g}^{2}N_{f}(\alpha H_{f}+2\beta H_{g})\right]\bigg\},
\end{eqnarray}
\begin{eqnarray}
\Xi_{24} & = & \frac{1}{\Delta}\frac{4k^{8}a_{g}^{4}M_{g}^{2}N_{f}^{2}}{3\tilde{N}^{3}a_{f}^{2}}\bigg\{\mathcal{C}\left(\frac{\mathrm{d}\bar{\phi}}{\mathrm{d}t}\right)^{2}\tilde{a}^{3}a_{g}^{2}g_{\phi\phi}\left(\alpha F_{24}N_{g}-\beta F_{14}N_{f}\right)\left(\alpha a_{f}^{3}H_{f}M_{f}^{2}N_{g}-\beta a_{g}^{3}H_{g}M_{g}^{2}N_{f}\right)\nonumber \\
&  & +\tilde{N}^{3}a_{f}^{3}H_{f}H_{g}M_{f}^{2}M_{g}^{2}\left[2F_{14}k^{2}a_{f}^{3}H_{f}M_{f}^{2}N_{g}-3\mathcal{C}a_{g}^{5}\left(F_{14}H_{f}N_{f}+F_{24}H_{g}N_{g}\right)\right]\bigg\},
\end{eqnarray}
\begin{eqnarray}
\Xi_{25} & = & \frac{1}{\Delta}\frac{4}{9\tilde{N}^{3}a_{f}^{2}}k^{12}a_{g}^{5}M_{g}^{4}N_{f}^{2}N_{g}\bigg\{\beta\mathcal{C}\left(\frac{\mathrm{d}\bar{\phi}}{\mathrm{d}t}\right)^{2}\tilde{a}^{3}a_{g}^{2}g_{\phi\phi}N_{f}\left(\beta a_{g}^{3}H_{g}M_{g}^{2}N_{f}-\alpha a_{f}^{3}H_{f}M_{f}^{2}N_{g}\right)\nonumber \\
&  & +\tilde{N}^{3}a_{f}^{3}H_{f}^{2}H_{g}M_{f}^{2}M_{g}^{2}\left(2k^{2}a_{f}^{3}M_{f}^{2}N_{g}-3\mathcal{C}a_{g}^{5}N_{f}\right)\bigg\},
\end{eqnarray}
\begin{eqnarray}
\Xi_{26} & = & \frac{1}{\Delta}\frac{4}{3\tilde{N}^{3}a_{f}^{2}}k^{8}a_{g}^{4}M_{g}^{2}N_{f}^{2}\bigg\{\mathcal{C}\left(\frac{\mathrm{d}\bar{\phi}}{\mathrm{d}t}\right)^{2}\tilde{a}^{3}a_{g}^{2}g_{\phi\phi}\left(\alpha F_{26}N_{g}-\beta F_{16}N_{f}\right)\left(\alpha a_{f}^{3}H_{f}M_{f}^{2}N_{g}-\beta a_{g}^{3}H_{g}M_{g}^{2}N_{f}\right)\nonumber \\
&  & +\tilde{N}^{3}a_{f}^{3}H_{f}H_{g}M_{f}^{2}M_{g}^{2}\left[2F_{16}k^{2}a_{f}^{3}H_{f}M_{f}^{2}N_{g}-3\mathcal{C}a_{g}^{5}\left(F_{16}H_{f}N_{f}+F_{26}H_{g}N_{g}\right)\right]\bigg\},
\end{eqnarray}
\begin{eqnarray}
\Xi_{34} & = & -\frac{1}{\Delta}\frac{8}{\tilde{N}^{3}}k^{6}a_{f}a_{g}^{3}M_{f}^{2}N_{f}N_{g}\bigg\{ F_{24}\tilde{N}^{3}a_{g}H_{f}H_{g}^{2}M_{g}^{2}\left[a_{f}^{3}M_{f}^{2}N_{g}\left(3\mathcal{C}a_{g}^{2}-2k^{2}M_{g}^{2}\right)+3\mathcal{C}a_{g}^{5}M_{g}^{2}N_{f}\right]\nonumber \\
&  & -\mathcal{C}\left(\frac{\mathrm{d}\bar{\phi}}{\mathrm{d}t}\right)^{2}\tilde{a}^{3}g_{\phi\phi}\left(\alpha F_{24}N_{g}-\beta F_{14}N_{f}\right)\left(\alpha a_{f}^{3}H_{f}M_{f}^{2}N_{g}-\beta a_{g}^{3}H_{g}M_{g}^{2}N_{f}\right)\bigg\},
\end{eqnarray}
\begin{equation}
\Xi_{35}=-\frac{1}{\Delta}\frac{8}{3\tilde{N}^{3}}\beta\mathcal{C}\left(\frac{\mathrm{d}\bar{\phi}}{\mathrm{d}t}\right)^{2}k^{10}\tilde{a}^{3}a_{f}a_{g}^{4}g_{\phi\phi}M_{f}^{2}M_{g}^{2}N_{f}^{2}N_{g}^{2}\left(\alpha a_{f}^{3}H_{f}M_{f}^{2}N_{g}-\beta a_{g}^{3}H_{g}M_{g}^{2}N_{f}\right),
\end{equation}
\begin{eqnarray}
\Xi_{36} & = & -\frac{1}{\Delta}\frac{8}{\tilde{N}^{3}}k^{6}a_{f}a_{g}^{3}M_{f}^{2}N_{f}N_{g}\bigg\{ F_{26}\tilde{N}^{3}a_{g}H_{f}H_{g}^{2}M_{g}^{2}\left[a_{f}^{3}M_{f}^{2}N_{g}\left(3\mathcal{C}a_{g}^{2}-2k^{2}M_{g}^{2}\right)+3\mathcal{C}a_{g}^{5}M_{g}^{2}N_{f}\right]\nonumber \\
&  & -\mathcal{C}\left(\frac{\mathrm{d}\bar{\phi}}{\mathrm{d}t}\right)^{2}\tilde{a}^{3}g_{\phi\phi}\left(\alpha F_{26}N_{g}-\beta F_{16}N_{f}\right)\left(\alpha a_{f}^{3}H_{f}M_{f}^{2}N_{g}-\beta a_{g}^{3}H_{g}M_{g}^{2}N_{f}\right)\bigg\},
\end{eqnarray}
\begin{eqnarray}
\Xi_{44} & = & -\frac{1}{\Delta}\frac{4k^{6}a_{g}^{3}N_{f}^{2}N_{g}}{\tilde{N}^{3}a_{f}^{2}}\bigg\{\tilde{N}^{3}\bigg[a_{f}^{6}H_{f}^{2}M_{f}^{4}\left(\mathcal{C}F_{14}^{2}-4k^{2}M_{44}a_{g}H_{g}^{2}M_{g}^{4}N_{g}+6\mathcal{C}M_{44}a_{g}^{3}H_{g}^{2}M_{g}^{2}N_{g}\right)\nonumber \\
&  & +2\mathcal{C}a_{f}^{3}a_{g}^{3}H_{f}H_{g}M_{f}^{2}M_{g}^{2}\left(3M_{44}a_{g}^{3}H_{f}H_{g}M_{g}^{2}N_{f}-F_{14}F_{24}\right)+\mathcal{C}F_{24}^{2}a_{g}^{6}H_{g}^{2}M_{g}^{4}\bigg]\nonumber \\
&  & -2\mathcal{C}\left(\frac{\mathrm{d}\bar{\phi}}{\mathrm{d}t}\right)^{2}M_{44}\tilde{a}^{3}g_{\phi\phi}\left(\alpha a_{f}^{3}H_{f}M_{f}^{2}N_{g}-\beta a_{g}^{3}H_{g}M_{g}^{2}N_{f}\right)^{2}\bigg\}
\end{eqnarray}
\begin{eqnarray}
\Xi_{45} & = & \frac{1}{\Delta}\frac{4}{3\tilde{N}^{3}a_{f}^{2}}k^{10}a_{g}^{4}M_{g}^{2}N_{f}^{2}N_{g}^{2}\bigg\{\tilde{N}^{3}a_{f}^{3}H_{f}M_{f}^{2}\bigg[\mathcal{C}a_{g}^{3}H_{g}M_{g}^{2}\left(F_{24}-6a_{g}^{3}H_{f}H_{g}M_{g}^{2}N_{f}\right)\nonumber \\
&  & -a_{f}^{3}H_{f}M_{f}^{2}\left(\mathcal{C}F_{14}-4k^{2}a_{g}H_{g}^{2}M_{g}^{4}N_{g}+6\mathcal{C}a_{g}^{3}H_{g}^{2}M_{g}^{2}N_{g}\right)\bigg]\nonumber \\
&  & +2\mathcal{C}\left(\frac{\mathrm{d}\bar{\phi}}{\mathrm{d}t}\right)^{2}\tilde{a}^{3}g_{\phi\phi}\left(\alpha a_{f}^{3}H_{f}M_{f}^{2}N_{g}-\beta a_{g}^{3}H_{g}M_{g}^{2}N_{f}\right)^{2}\bigg\},
\end{eqnarray}
\begin{eqnarray}
\Xi_{46} & = & \frac{1}{\Delta} \frac{4}{\tilde{N}^{3}a_{f}^{2}}k^{6}a_{g}^{3}N_{f}^{2}N_{g}\bigg\{\tilde{N}^{3}\bigg[\mathcal{C}a_{f}^{3}a_{g}^{3}H_{f}H_{g}M_{f}^{2}M_{g}^{2}\left(F_{14}F_{26}+F_{16}F_{24}-6M_{46}a_{g}^{3}H_{f}H_{g}M_{g}^{2}N_{f}\right)\nonumber \\
&  & -a_{f}^{6}H_{f}^{2}M_{f}^{4}\left(\mathcal{C}F_{14}F_{16}-4k^{2}M_{46}a_{g}H_{g}^{2}M_{g}^{4}N_{g}+6\mathcal{C}M_{46}a_{g}^{3}H_{g}^{2}M_{g}^{2}N_{g}\right)-\mathcal{C}F_{24}F_{26}a_{g}^{6}H_{g}^{2}M_{g}^{4}\bigg]\nonumber \\
&  & +2\mathcal{C}\left(\frac{\mathrm{d}\bar{\phi}}{\mathrm{d}t}\right)^{2}M_{46}\tilde{a}^{3}g_{\phi\phi}\left(\alpha a_{f}^{3}H_{f}M_{f}^{2}N_{g}-\beta a_{g}^{3}H_{g}M_{g}^{2}N_{f}\right)^{2}\bigg\},
\end{eqnarray}
\begin{eqnarray}
\Xi_{55} & = & -\frac{1}{\Delta}\frac{4k^{6}a_{g}^{3}N_{f}^{2}N_{g}}{9\tilde{N}^{3}a_{f}^{2}}\bigg\{\tilde{N}^{3}a_{f}^{3}a_{g}H_{f}^{2}M_{f}^{2}M_{g}^{2}\bigg[a_{f}^{3}M_{f}^{2}N_{g}\left(\mathcal{C}k^{8}a_{g}M_{g}^{2}N_{g}+H_{g}^{2}\left(54\mathcal{C}M_{55}a_{g}^{2}-36k^{2}M_{55}M_{g}^{2}\right)\right)\nonumber \\
&  & +54\mathcal{C}M_{55}a_{g}^{5}H_{g}^{2}M_{g}^{2}N_{f}\bigg]-18\mathcal{C}\left(\frac{\mathrm{d}\bar{\phi}}{\mathrm{d}t}\right)^{2}M_{55}\tilde{a}^{3}g_{\Phi\Phi}\left(\alpha a_{f}^{3}H_{f}M_{f}^{2}N_{g}-\beta a_{g}^{3}H_{g}M_{g}^{2}N_{f}\right)^{2}\bigg\}.
\end{eqnarray}
\begin{equation}
\Xi_{56}=-\frac{1}{\Delta}\frac{4}{3}\mathcal{C}k^{10}a_{f}a_{g}^{4}H_{f}M_{f}^{2}M_{g}^{2}N_{f}^{2}N_{g}^{2}\left(F_{16}a_{f}^{3}H_{f}M_{f}^{2}-F_{26}a_{g}^{3}H_{g}M_{g}^{2}\right),
\end{equation}
\begin{eqnarray}
\Xi_{66} & = & -\frac{1}{\Delta}\frac{4k^{6}a_{g}^{3}N_{f}^{2}N_{g}}{\tilde{N}^{3}a_{f}^{2}}\bigg\{\tilde{N}^{3}\bigg[a_{f}^{6}H_{f}^{2}M_{f}^{4}\left(\mathcal{C}F_{16}^{2}-4k^{2}M_{66}a_{g}H_{g}^{2}M_{g}^{4}N_{g}+6\mathcal{C}M_{66}a_{g}^{3}H_{g}^{2}M_{g}^{2}N_{g}\right)\nonumber \\
&  & +2\mathcal{C}a_{f}^{3}a_{g}^{3}H_{f}H_{g}M_{f}^{2}M_{g}^{2}\left(3M_{66}a_{g}^{3}H_{f}H_{g}M_{g}^{2}N_{f}-F_{15}F_{25}\right)+\mathcal{C}F_{26}^{2}a_{g}^{6}H_{g}^{2}M_{g}^{4}\bigg]\nonumber \\
&  & -2\mathcal{C}\left(\frac{\mathrm{d}\bar{\phi}}{\mathrm{d}t}\right)^{2}M_{66}\tilde{a}^{3}g_{\phi\phi}\left(\alpha a_{f}^{3}H_{f}M_{f}^{2}N_{g}-\beta a_{g}^{3}H_{g}M_{g}^{2}N_{f}\right)^{2}\bigg\}.
\end{eqnarray}
In the above
	\begin{equation}
		g_{\phi\phi} = \frac{1}{2}\left(P_{,\tilde{X}}+2\tilde{X}P_{,\tilde{X}\tilde{X}}\right),
	\end{equation}
$\mathcal{C}$ and $\mathcal{M}$ are given in (\ref{Ccal}) and (\ref{Mcal}), respectively, and
	\begin{equation}
	F_{14}=a_{g}N_{g}\left[2k^{2}M_{g}^{2}+3\alpha^{2}\tilde{a}^{2}\left(P-2\tilde{X}P_{,\tilde{X}}\right)+3a_{g}^{2}(b_{1}-M_{g}^{2}\Lambda_{g}+3M_{g}^{2}H_{g}^{2})\right],
	\end{equation}
	\begin{equation}
	F_{16}=3N_{g}\left[\alpha\beta\tilde{a}^{2}a_{f}\left(P-2\tilde{X}P_{,\tilde{X}}\right)+a_{g}^{3}b_{2}\right],
	\end{equation}
	\begin{equation}
	F_{24}=\frac{a_{g}N_{f}}{a_{f}N_{g}}F_{16},
	\end{equation}
	\begin{equation}
	F_{26}=a_{f}N_{f}\left[2k^{2}M_{f}^{2}+3\beta^{2}\tilde{a}^{2}\left(P-2\tilde{X}P_{,\tilde{X}}\right)+3a_{f}^{2}\left(-M_{f}^{2}\Lambda_{f}+3M_{f}^{2}H_{f}^{2}\right)\right]+3a_{g}^{3}b_{3}N_{g},
	\end{equation}
	\begin{eqnarray}
	M_{44} & = & 2k^{2}a_{g}M_{g}^{2}N_{g}+3a_{g}^{2}\Bigg\{3\left(m^{2}a_{f}M_{g}^{2}(c_{1}N_{g}+2c_{2}N_{f})+\alpha^{2}\tilde{a}\tilde{N}P\right)\nonumber \\
	&  & +a_{g}\left[N_{g}M_{g}^{2}\left(3m^{2}c_{0}-3\Lambda_{g}+9H_{g}^{2}+3\frac{2}{N_{g}}\frac{\mathrm{d}H_{g}}{\mathrm{d}t}\right)+N_{g}\mathcal{M}^{2}+3m^{2}c_{1}M_{g}^{2}N_{f}\right]\Bigg\},
	\end{eqnarray}
	\begin{eqnarray}
	M_{46} & = & 3a_{g}\bigg[3a_{f}\left(m^{2}a_{g}M_{g}^{2}(c_{1}N_{g}+2c_{2}N_{f})+\alpha\beta\tilde{a}\tilde{N}P\right)\nonumber \\
	&  & +6m^{2}a_{f}^{2}M_{g}^{2}(c_{2}N_{g}+3c_{3}N_{f})-a_{g}^{2}N_{g}\mathcal{M}^{2}\bigg],
	\end{eqnarray}
	\begin{equation}
	M_{55}=\frac{1}{18}k^{6}a_{g}M_{g}^{2}N_{g}+\frac{1}{6}k^{2}a_{g}^{3}N_{g}\mathcal{M}^{2},
	\end{equation}
	\begin{eqnarray}
	M_{66} & = & 2k^{2}a_{f}M_{f}^{2}N_{f}+3a_{g}^{3}N_{g}\mathcal{M}^{2}+9a_{f}{}^{2}\left(2m^{2}a_{g}M_{g}^{2}(c_{2}N_{g}+3c_{3}N_{f})+\beta^{2}\tilde{a}\tilde{N}P\right)\nonumber \\
	&  & +9a_{f}^{3}\left[6m^{2}c_{3}M_{g}^{2}N_{g}+N_{f}M_{g}^{2}\left(24m^{2}c_{4}-\Lambda_{f}\right)+N_{f}M_{f}^{2}\left(3H_{f}^{2}+\frac{2}{N_{f}}\frac{\mathrm{d}H_{f}}{\mathrm{d}t}\right)\right].
	\end{eqnarray}


\bibliographystyle{JHEPmodplain}
\bibliography{D:/Dropbox/BIBTEX/Articles}

\providecommand{\href}[2]{#2}\begingroup\raggedright\begin{thebibliography}{10}

\bibitem{Ade:2015hxq}
{\bf Planck} Collaboration, P.~A.~R. Ade et~al., {\it {Planck 2015 results.
  XVI. Isotropy and statistics of the CMB}},
  \href{http://arxiv.org/abs/1506.07135}{{\tt arXiv:1506.07135}}.

\bibitem{Ade:2015xua}
{\bf Planck} Collaboration, P.~A.~R. Ade et~al., {\it {Planck 2015 results.
  XIII. Cosmological parameters}},  \href{http://arxiv.org/abs/1502.01589}{{\tt
  arXiv:1502.01589}}.

\bibitem{Ade:2015lrj}
{\bf Planck} Collaboration, P.~Ade et~al., {\it {Planck 2015. XX. Constraints
  on inflation}},  \href{http://arxiv.org/abs/1502.02114}{{\tt
  arXiv:1502.02114}}.

\bibitem{Amendola:2012ys}
{\bf Euclid Theory Working Group} Collaboration, L.~Amendola et~al., {\it
  {Cosmology and fundamental physics with the Euclid satellite}},  {\em Living
  Rev. Rel.} {\bf 16} (2013) 6, [\href{http://arxiv.org/abs/1206.1225}{{\tt
  arXiv:1206.1225}}].

\bibitem{Weinberg:1988cp}
S.~Weinberg, {\it {The Cosmological Constant Problem}},  {\em Rev. Mod. Phys.}
  {\bf 61} (1989) 1--23.

\bibitem{Horndeski:1974wa}
G.~W. Horndeski, {\it {Second-order scalar-tensor field equations in a
  four-dimensional space}},  {\em Int.J.Theor.Phys.} {\bf 10} (1974) 363--384.

\bibitem{Nicolis:2008in}
A.~Nicolis, R.~Rattazzi, and E.~Trincherini, {\it {The Galileon as a local
  modification of gravity}},  {\em Phys.Rev.} {\bf D79} (2009) 064036,
  [\href{http://arxiv.org/abs/0811.2197}{{\tt arXiv:0811.2197}}].

\bibitem{Deffayet:2009wt}
C.~Deffayet, G.~Esposito-Farese, and A.~Vikman, {\it {Covariant Galileon}},
  {\em Phys.Rev.} {\bf D79} (2009) 084003,
  [\href{http://arxiv.org/abs/0901.1314}{{\tt arXiv:0901.1314}}].

\bibitem{Deffayet:2009mn}
C.~Deffayet, S.~Deser, and G.~Esposito-Farese, {\it {Generalized Galileons: All
  scalar models whose curved background extensions maintain second-order field
  equations and stress-tensors}},  {\em Phys.Rev.} {\bf D80} (2009) 064015,
  [\href{http://arxiv.org/abs/0906.1967}{{\tt arXiv:0906.1967}}].

\bibitem{Deffayet:2011gz}
C.~Deffayet, X.~Gao, D.~Steer, and G.~Zahariade, {\it {From k-essence to
  generalised Galileons}},  {\em Phys.Rev.} {\bf D84} (2011) 064039,
  [\href{http://arxiv.org/abs/1103.3260}{{\tt arXiv:1103.3260}}].

\bibitem{DeFelice:2010pv}
A.~De~Felice and S.~Tsujikawa, {\it {Cosmology of a covariant Galileon field}},
   {\em Phys. Rev. Lett.} {\bf 105} (2010) 111301,
  [\href{http://arxiv.org/abs/1007.2700}{{\tt arXiv:1007.2700}}].

\bibitem{Kobayashi:2010cm}
T.~Kobayashi, M.~Yamaguchi, and J.~Yokoyama, {\it {G-inflation: Inflation
  driven by the Galileon field}},  {\em Phys.Rev.Lett.} {\bf 105} (2010)
  231302, [\href{http://arxiv.org/abs/1008.0603}{{\tt arXiv:1008.0603}}].

\bibitem{Kobayashi:2011nu}
T.~Kobayashi, M.~Yamaguchi, and J.~Yokoyama, {\it {Generalized G-inflation:
  Inflation with the most general second-order field equations}},  {\em
  Prog.Theor.Phys.} {\bf 126} (2011) 511--529,
  [\href{http://arxiv.org/abs/1105.5723}{{\tt arXiv:1105.5723}}].

\bibitem{Gao:2011qe}
X.~Gao and D.~A. Steer, {\it {Inflation and primordial non-Gaussianities of
  'generalized Galileons'}},  {\em JCAP} {\bf 1112} (2011) 019,
  [\href{http://arxiv.org/abs/1107.2642}{{\tt arXiv:1107.2642}}].

\bibitem{Gao:2011mz}
X.~Gao, {\it {Conserved cosmological perturbation in Galileon models}},  {\em
  JCAP} {\bf 1110} (2011) 021, [\href{http://arxiv.org/abs/1106.0292}{{\tt
  arXiv:1106.0292}}].

\bibitem{Gao:2011vs}
X.~Gao, T.~Kobayashi, M.~Yamaguchi, and J.~Yokoyama, {\it {Primordial
  non-Gaussianities of gravitational waves in the most general single-field
  inflation model}},  {\em Phys.Rev.Lett.} {\bf 107} (2011) 211301,
  [\href{http://arxiv.org/abs/1108.3513}{{\tt arXiv:1108.3513}}].

\bibitem{DeFelice:2011uc}
A.~De~Felice and S.~Tsujikawa, {\it {Inflationary non-Gaussianities in the most
  general second-order scalar-tensor theories}},  {\em Phys.Rev.} {\bf D84}
  (2011) 083504, [\href{http://arxiv.org/abs/1107.3917}{{\tt
  arXiv:1107.3917}}].

\bibitem{deRham:2011by}
C.~de~Rham and L.~Heisenberg, {\it {Cosmology of the Galileon from Massive
  Gravity}},  {\em Phys.Rev.} {\bf D84} (2011) 043503,
  [\href{http://arxiv.org/abs/1106.3312}{{\tt arXiv:1106.3312}}].

\bibitem{Heisenberg:2014kea}
L.~Heisenberg, R.~Kimura, and K.~Yamamoto, {\it {Cosmology of the proxy theory
  to massive gravity}},  {\em Phys.Rev.} {\bf D89} (2014), no.~10 103008,
  [\href{http://arxiv.org/abs/1403.2049}{{\tt arXiv:1403.2049}}].

\bibitem{Horndeski:1976gi}
G.~W. Horndeski, {\it {Conservation of Charge and the Einstein-Maxwell Field
  Equations}},  {\em J. Math. Phys.} {\bf 17} (1976) 1980--1987.

\bibitem{EspositoFarese:2009aj}
G.~Esposito-Farese, C.~Pitrou, and J.-P. Uzan, {\it {Vector theories in
  cosmology}},  {\em Phys. Rev.} {\bf D81} (2010) 063519,
  [\href{http://arxiv.org/abs/0912.0481}{{\tt arXiv:0912.0481}}].

\bibitem{Jimenez:2009py}
J.~Beltran~Jimenez, R.~Lazkoz, and A.~L. Maroto, {\it {Cosmic vector for dark
  energy: Constraints from supernovae, cosmic microwave background, and baryon
  acoustic oscillations}},  {\em Phys. Rev.} {\bf D80} (2009) 023004,
  [\href{http://arxiv.org/abs/0904.0433}{{\tt arXiv:0904.0433}}].

\bibitem{BeltranJimenez:2013fca}
J.~B. Jimenez, A.~L. Delvas~Froes, and D.~F. Mota, {\it {Screening Vector Field
  Modifications of General Relativity}},  {\em Phys. Lett.} {\bf B725} (2013)
  212--217, [\href{http://arxiv.org/abs/1212.1923}{{\tt arXiv:1212.1923}}].

\bibitem{Jimenez:2013qsa}
J.~B. Jiménez, R.~Durrer, L.~Heisenberg, and M.~Thorsrud, {\it {Stability of
  Horndeski vector-tensor interactions}},  {\em JCAP} {\bf 1310} (2013) 064,
  [\href{http://arxiv.org/abs/1308.1867}{{\tt arXiv:1308.1867}}].

\bibitem{Jimenez:2014rna}
J.~Beltrán~Jiménez and T.~S. Koivisto, {\it {Extended Gauss-Bonnet gravities
  in Weyl geometry}},  {\em Class. Quant. Grav.} {\bf 31} (2014) 135002,
  [\href{http://arxiv.org/abs/1402.1846}{{\tt arXiv:1402.1846}}].

\bibitem{Heisenberg:2014rta}
L.~Heisenberg, {\it {Generalization of the Proca Action}},  {\em JCAP} {\bf
  1405} (2014) 015, [\href{http://arxiv.org/abs/1402.7026}{{\tt
  arXiv:1402.7026}}].

\bibitem{Tasinato:2014eka}
G.~Tasinato, {\it {Cosmic Acceleration from Abelian Symmetry Breaking}},  {\em
  JHEP} {\bf 04} (2014) 067, [\href{http://arxiv.org/abs/1402.6450}{{\tt
  arXiv:1402.6450}}].

\bibitem{Allys:2015sht}
E.~Allys, P.~Peter, and Y.~Rodriguez, {\it {Generalized Proca action for an
  Abelian vector field}},  \href{http://arxiv.org/abs/1511.03101}{{\tt
  arXiv:1511.03101}}.

\bibitem{Fierz:1939ix}
M.~Fierz and W.~Pauli, {\it {On relativistic wave equations for particles of
  arbitrary spin in an electromagnetic field}},  {\em Proc.Roy.Soc.Lond.} {\bf
  A173} (1939) 211--232.

\bibitem{vanDam:1970vg}
H.~van Dam and M.~Veltman, {\it {Massive and massless Yang-Mills and
  gravitational fields}},  {\em Nucl.Phys.} {\bf B22} (1970) 397--411.

\bibitem{Zakharov:1970cc}
V.~Zakharov, {\it {Linearized gravitation theory and the graviton mass}},  {\em
  JETP Lett.} {\bf 12} (1970) 312.

\bibitem{Vainshtein:1972sx}
A.~Vainshtein, {\it {To the problem of nonvanishing gravitation mass}},  {\em
  Phys.Lett.} {\bf B39} (1972) 393--394.

\bibitem{Boulware:1973my}
D.~G. Boulware and S.~Deser, {\it {Can gravitation have a finite range?}},
  {\em Phys. Rev.} {\bf D6} (1972) 3368--3382.

\bibitem{deRham:2010ik}
C.~de~Rham and G.~Gabadadze, {\it {Generalization of the Fierz-Pauli Action}},
  {\em Phys.Rev.} {\bf D82} (2010) 044020,
  [\href{http://arxiv.org/abs/1007.0443}{{\tt arXiv:1007.0443}}].

\bibitem{deRham:2010kj}
C.~de~Rham, G.~Gabadadze, and A.~J. Tolley, {\it {Resummation of Massive
  Gravity}},  {\em Phys.Rev.Lett.} {\bf 106} (2011) 231101,
  [\href{http://arxiv.org/abs/1011.1232}{{\tt arXiv:1011.1232}}].

\bibitem{Hassan:2011vm}
S.~Hassan and R.~A. Rosen, {\it {On Non-Linear Actions for Massive Gravity}},
  {\em JHEP} {\bf 1107} (2011) 009, [\href{http://arxiv.org/abs/1103.6055}{{\tt
  arXiv:1103.6055}}].

\bibitem{Hassan:2011hr}
S.~Hassan and R.~A. Rosen, {\it {Resolving the Ghost Problem in non-Linear
  Massive Gravity}},  {\em Phys.Rev.Lett.} {\bf 108} (2012) 041101,
  [\href{http://arxiv.org/abs/1106.3344}{{\tt arXiv:1106.3344}}].

\bibitem{Hassan:2011zd}
S.~Hassan and R.~A. Rosen, {\it {Bimetric Gravity from Ghost-free Massive
  Gravity}},  {\em JHEP} {\bf 1202} (2012) 126,
  [\href{http://arxiv.org/abs/1109.3515}{{\tt arXiv:1109.3515}}].

\bibitem{deRham:2012ew}
C.~de~Rham, G.~Gabadadze, L.~Heisenberg, and D.~Pirtskhalava, {\it
  {Nonrenormalization and naturalness in a class of scalar-tensor theories}},
  {\em Phys. Rev.} {\bf D87} (2013), no.~8 085017,
  [\href{http://arxiv.org/abs/1212.4128}{{\tt arXiv:1212.4128}}].

\bibitem{deRham:2013qqa}
C.~de~Rham, L.~Heisenberg, and R.~H. Ribeiro, {\it {Quantum Corrections in
  Massive Gravity}},  {\em Phys. Rev.} {\bf D88} (2013) 084058,
  [\href{http://arxiv.org/abs/1307.7169}{{\tt arXiv:1307.7169}}].

\bibitem{deRham:2014naa}
C.~de~Rham, L.~Heisenberg, and R.~H. Ribeiro, {\it {On couplings to matter in
  massive (bi-)gravity}},  {\em Class. Quant. Grav.} {\bf 32} (2015) 035022,
  [\href{http://arxiv.org/abs/1408.1678}{{\tt arXiv:1408.1678}}].

\bibitem{Noller:2014sta}
J.~Noller and S.~Melville, {\it {The coupling to matter in Massive, Bi- and
  Multi-Gravity}},  {\em JCAP} {\bf 1501} (2015) 003,
  [\href{http://arxiv.org/abs/1408.5131}{{\tt arXiv:1408.5131}}].

\bibitem{Heisenberg:2014rka}
L.~Heisenberg, {\it {Quantum corrections in massive bigravity and new effective
  composite metrics}},  {\em Class. Quant. Grav.} {\bf 32} (2015), no.~10
  105011, [\href{http://arxiv.org/abs/1410.4239}{{\tt arXiv:1410.4239}}].

\bibitem{deRham:2014fha}
C.~de~Rham, L.~Heisenberg, and R.~H. Ribeiro, {\it {Ghosts \& Matter Couplings
  in Massive (bi-\&multi-)Gravity}},  {\em Phys.Rev.} {\bf D90} (2014) 124042,
  [\href{http://arxiv.org/abs/1409.3834}{{\tt arXiv:1409.3834}}].

\bibitem{Huang:2015yga}
Q.-G. Huang, R.~H. Ribeiro, Y.-H. Xing, K.-C. Zhang, and S.-Y. Zhou, {\it {On
  the uniqueness of the non-minimal matter coupling in massive gravity and
  bigravity}},  {\em Phys. Lett.} {\bf B748} (2015) 356--360,
  [\href{http://arxiv.org/abs/1505.02616}{{\tt arXiv:1505.02616}}].

\bibitem{Heisenberg:2015iqa}
L.~Heisenberg, {\it {More on effective composite metrics}},  {\em Phys. Rev.}
  {\bf D92} (2015) 023525, [\href{http://arxiv.org/abs/1505.02966}{{\tt
  arXiv:1505.02966}}].

\bibitem{Melville:2015dba}
S.~Melville and J.~Noller, {\it {Generalised matter couplings in massive
  bigravity}},  \href{http://arxiv.org/abs/1511.01485}{{\tt arXiv:1511.01485}}.

\bibitem{Hinterbichler:2015yaa}
K.~Hinterbichler and R.~A. Rosen, {\it {Note on ghost-free matter couplings in
  massive gravity and multigravity}},  {\em Phys. Rev.} {\bf D92} (2015), no.~2
  024030, [\href{http://arxiv.org/abs/1503.06796}{{\tt arXiv:1503.06796}}].

\bibitem{deRham:2015cha}
C.~de~Rham and A.~J. Tolley, {\it {Vielbein to the rescue? Breaking the
  symmetric vielbein condition in massive gravity and multigravity}},  {\em
  Phys. Rev.} {\bf D92} (2015), no.~2 024024,
  [\href{http://arxiv.org/abs/1505.01450}{{\tt arXiv:1505.01450}}].

\bibitem{DeFelice:2015yha}
A.~De~Felice, A.~E. Gümrükçüoğlu, L.~Heisenberg, and S.~Mukohyama, {\it
  {Matter coupling in partially constrained vielbein formulation of massive
  gravity}},  \href{http://arxiv.org/abs/1509.05978}{{\tt arXiv:1509.05978}}.

\bibitem{Enander:2014xga}
J.~Enander, A.~R. Solomon, Y.~Akrami, and E.~Mortsell, {\it {Cosmic expansion
  histories in massive bigravity with symmetric matter coupling}},  {\em JCAP}
  {\bf 1501} (2015) 006, [\href{http://arxiv.org/abs/1409.2860}{{\tt
  arXiv:1409.2860}}].

\bibitem{Gumrukcuoglu:2014xba}
A.~Emir~Gümrükçüoğlu, L.~Heisenberg, and S.~Mukohyama, {\it {Cosmological
  perturbations in massive gravity with doubly coupled matter}},  {\em JCAP}
  {\bf 1502} (2015) 022, [\href{http://arxiv.org/abs/1409.7260}{{\tt
  arXiv:1409.7260}}].

\bibitem{Solomon:2014iwa}
A.~R. Solomon, J.~Enander, Y.~Akrami, T.~S. Koivisto, F.~Könnig, and
  E.~Mörtsell, {\it {Cosmological viability of massive gravity with
  generalized matter coupling}},  {\em JCAP} {\bf 1504} (2015), no.~04 027,
  [\href{http://arxiv.org/abs/1409.8300}{{\tt arXiv:1409.8300}}].

\bibitem{Gao:2014xaa}
X.~Gao and D.~Yoshida, {\it {Coupling between Galileon and massive gravity with
  composite metrics}},  {\em Phys. Rev.} {\bf D92} (2015), no.~4 044057,
  [\href{http://arxiv.org/abs/1412.8471}{{\tt arXiv:1412.8471}}].

\bibitem{Comelli:2015pua}
D.~Comelli, M.~Crisostomi, K.~Koyama, L.~Pilo, and G.~Tasinato, {\it {Cosmology
  of bigravity with doubly coupled matter}},  {\em JCAP} {\bf 1504} (2015) 026,
  [\href{http://arxiv.org/abs/1501.00864}{{\tt arXiv:1501.00864}}].

\bibitem{Gumrukcuoglu:2015nua}
A.~E. Gumrukcuoglu, L.~Heisenberg, S.~Mukohyama, and N.~Tanahashi, {\it
  {Cosmology in bimetric theory with an effective composite coupling to
  matter}},  {\em JCAP} {\bf 1504} (2015), no.~04 008,
  [\href{http://arxiv.org/abs/1501.02790}{{\tt arXiv:1501.02790}}].

\bibitem{Lagos:2015sya}
M.~Lagos and J.~Noller, {\it {New massive bigravity cosmologies with double
  matter coupling}},  \href{http://arxiv.org/abs/1508.05864}{{\tt
  arXiv:1508.05864}}.

\bibitem{Heisenberg:2015wja}
L.~Heisenberg, {\it {Non-minimal derivative couplings of the composite
  metric}},  {\em JCAP} {\bf 1511} (2015), no.~11 005,
  [\href{http://arxiv.org/abs/1506.00580}{{\tt arXiv:1506.00580}}].

\bibitem{Heisenberg:2016spl}
L.~Heisenberg and A.~Refregier, {\it {Cosmology in massive gravity with
  effective composite metric}},  \href{http://arxiv.org/abs/1604.07306}{{\tt
  arXiv:1604.07306}}.

\bibitem{Heisenberg:2016dkj}
L.~Heisenberg and A.~Refregier, {\it {Cosmology in doubly coupled massive
  gravity: constraints from SNIa, BAO and CMB}},
  \href{http://arxiv.org/abs/1604.07680}{{\tt arXiv:1604.07680}}.

\bibitem{Blanchet:2015sra}
L.~Blanchet and L.~Heisenberg, {\it {Dark Matter via Massive (bi-)Gravity}},
  {\em Phys.Rev.} {\bf D91} (2015) 103518,
  [\href{http://arxiv.org/abs/1504.00870}{{\tt arXiv:1504.00870}}].

\bibitem{Blanchet:2015bia}
L.~Blanchet and L.~Heisenberg, {\it {Dipolar Dark Matter with Massive
  Bigravity}},  {\em JCAP} {\bf 1512} (2015), no.~12 026,
  [\href{http://arxiv.org/abs/1505.05146}{{\tt arXiv:1505.05146}}].

\bibitem{Bernard:2015gwa}
L.~Bernard, L.~Blanchet, and L.~Heisenberg, {\it {Bimetric gravity and dark
  matter}},  in {\em {Proceedings, 50th Recontres de Moriond Gravitaion : 100
  years after GR}}, pp.~43--52, 2015.
\newblock \href{http://arxiv.org/abs/1507.02802}{{\tt arXiv:1507.02802}}.

\bibitem{Gao:2016vtt}
X.~Gao and L.~Heisenberg, {\it {Derivative couplings in massive bigravity}},
  {\em JCAP} {\bf 1603} (2016), no.~03 043,
  [\href{http://arxiv.org/abs/1601.02180}{{\tt arXiv:1601.02180}}].

\bibitem{Brax:2016ssf}
P.~Brax, A.-C. Davis, and J.~Noller, {\it {Dark Energy and Doubly Coupled
  Bigravity}},  \href{http://arxiv.org/abs/1606.05590}{{\tt arXiv:1606.05590}}.

\end{thebibliography}\endgroup

\end{document}